\documentclass[12pt]{article}
\usepackage{graphicx}
\usepackage{rotating}
\usepackage{amsmath,amssymb}
\begin{document}
\begin{center}
\Large\bf
{Energy dependence of Ti/Fe ratio  in the Galactic cosmic rays measured by the ATIC-2 experiment}
\end{center}
V.I.~Zatsepin$^1$, 
 A.D.~Panov$^1$,
N.V.~Sokolskaya$^1$,
 J.H.~Adams, Jr.$^2$, 
H.S.~Ahn$^3$, 
G.L.~Bashindzhagyan$^1$,  
J.~Chang$^4$, 
M.~Christl$^2$, 
A.R.~Fazely$^5$,  
 T.G.~Guzik$^6$,
 J.B.~Isbert$^6$, 
K.C.~Kim$^3$,
E.N.~Kouznetsov$^2$, 
M.I.~Panasyuk$^1$,  
E.S.~Seo$^3$,  
J.~Watts$^2$, 
J.P.~Wefel$^6$,
J.~Wu$^3$\\

\noindent
1. Skobeltsyn Institute of Nuclear Physics, Moscow State University, Moscow, Russia\\
2. NASA Marshall Space Flight Center, Huntsville, AL, U.S.A.\\
3. University of Maryland, College Park, MD, USA\\
4. Purple Mountain Observatory, Nanjing, P.R. China\\
5. Southern University, Baton Rouge, LA, U.S.A\\
6. Louisiana State University, Baton Rouge, LA, U.S.A\\

 Titanium is a rare, secondary nucleus among Galactic cosmic rays.  Using the Silicon matrix  in the ATIC experiment, Titanium has been separated. The energy dependence of the Ti to Fe flux ratio in the energy region from 5 GeV per nucleon to about 500 GeV per nucleon is presented.
\section{Introduction}
The rare elements such as Li, Be, B, Sc, Ti, V are present in the Galactic cosmic rays, because they are produced in the interactions of primary nuclei with the interstallar medium. Therefore, the energy dependence of a secondary to primary nuclei ratio can be used to study the medium through which cosmic rays propagate, as well as the ability of the Galactic magnetic fields to confine particles of different energy. The most accurate measurements of the energy dependence of secondary to primary ratio  were performed more than 30 years ago onboard the HEAO-3 satellite by Engelmann et al. \cite{engelman}, over the energy range from 0.62 GeV per nucleon to 35 GeV per nucleon for all nuclei from berrilium to nickel. In this experiment, as well as in earlier experiments (see ref in \cite{engelman})   it was found  that secondary to primary ratios decrease with  increasing energy according to the Leaky-Box model with the escape length which depends on the particle magnetic rigidity $R$ as $R^{-0.6}$. This dependence means that high energy particles escape more rapidly and thus traverse less material in the Galaxy than particles of lower energy. It also means that the acceleration of the cosmic ray  particles  occurs before they begin to propagate. Conversely, if the acceleration process requires more time to reach higher energy, we would expect secondary to primary  ratios to be constant, or even icrease with energy. It is obvious, however, that $R^{-0.6}$ dependence could not extend to very high energy ($\sim 10^{15}$ eV), because it comes into conflict with the existing EAS data on anysotropy \cite{erlykin}.  Osborne and Ptuskin \cite{ptuskin} suggested that the Galactic magnetic irregularities are of the Kolmogorov type and  lead to the dependence $R^{-0.33}$ at the high energy. To describe the energy dependence of secondary to primary nuclei measured in \cite{engelman}, they included the process of weak reacceleration in the interstellar medium. Moreover, both the primary and secondary nuclei should encounter  supernova remnants during their propagation, and this  may lead to an even flatter energy dependance of their ratios \cite{berezhko}.

Consequently, measurements of energy dependence for secondary to primary ratio is a very important problem, especially above 35 GeV per nucleon, where accurate experimental data are lacking.
\section{The ATIC experiment} 
The ATIC spectrometer was built for measurements of energy spectra of elements from H to Fe in the energy region from 100 GeV to few tens TeV. ATIC had three long-duration flights around the South Pole: ATIC-1 in December 2000 -- January 2001, ATIC-2 in December 2002 -- January 2003 and ATIC-4 in December 2007 -- January 2008. The ATIC-1 flight was a test one; the difficulties we met  during its processing did not allow yet to obtain the reliable data. The data from the science flight ATIC-2 on abundant nuclei have been analized and published in a series of papers ( see \cite{panov1},\cite{wefel}, for example). The possibility of measuring the energy dependence of boron to carbon ratio was examined in \cite{panov2}. In the present paper, we show that ATIC-2 has also the capability to measure the energy dependence of the titanum to iron ratio. The ATIC-4 experiment is now in the first stage of processing.

ATIC is comprised of a fully active bismuth germanate (BGO) calorimeter, a carbon target with embedded scintillator hodoscopes, and a silicon matrix that is used as a charge detector in the experiment. The ATIC calorimeter is thin, that is it measures only a part of the particle energy. The total particle energy was determined with  the formula $E(Z)=E_{d}(Z)/K(E_{d},Z)$, where E is the particle energy, Z is its charge and $E_{d}(Z)$ is the energy  deposited by this particle in the calorimeter. The coefficients  $K(E_{d},Z)$ were determined with the simulation of cascade development using the FLUKA codes \cite{fluka}. The characteristics of the ATIC spectrometer are 
decribed in \cite{guzik}, \cite{zatsepin}.
\begin{figure}[t]
 \begin{center}
 \includegraphics[width=36pc]{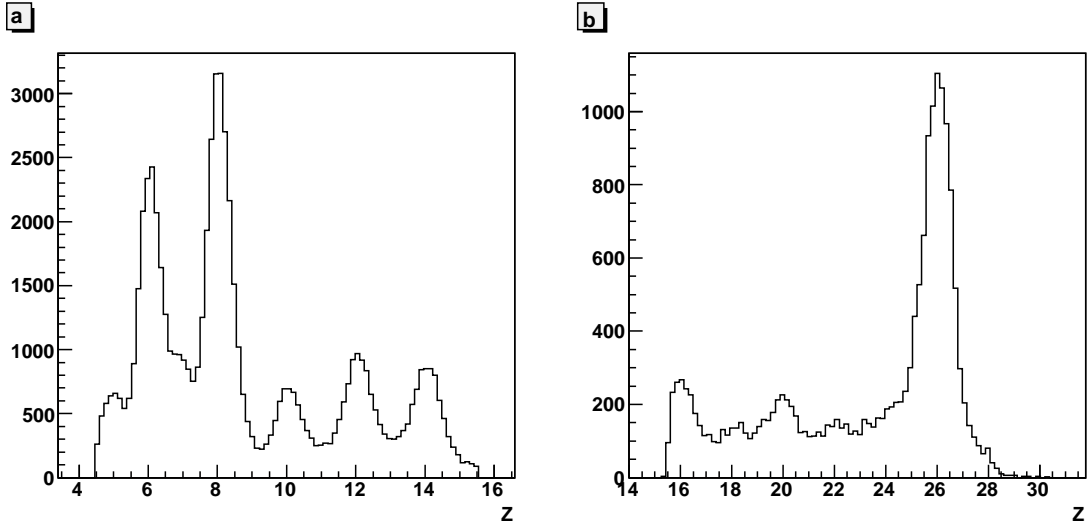}
  \end{center}
 \vspace{-1.0pc}
\caption{ Charge resolution in ATIC-2 at energy deposition above 50 GeV} 
\end{figure}
\begin{figure}[t]
 \begin{center}
 \includegraphics[width=36pc]{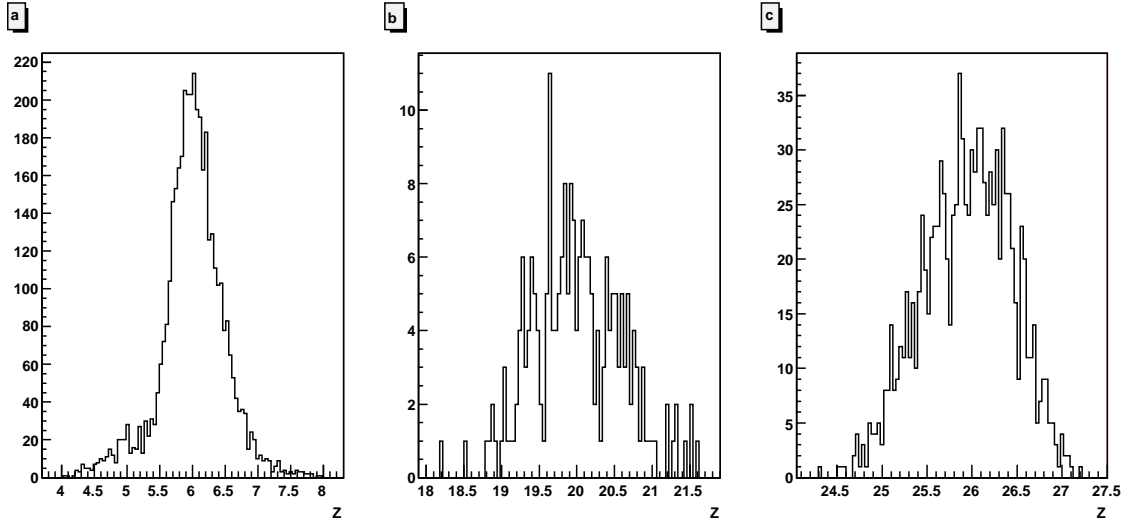}
  \end{center}
 \vspace{-1.0pc}
\caption{ Shape of charge lines in the double layer silicon matrix: a) carbon, b)calcium, c)iron } 
\end{figure}
\section {Charge measurements}
The trajectories of  nuclei were reconstructed using the data from the eightlayer BGO calorimeter. The intersection of the trajectory with the silicon matrix plane determines the coordinates of nucleus entering the silicon matrix, and the  uncertainty of these values for  each coordinate. The silicon pixel with the largest signal was found in  the area of confusion. This signal ($A$)  was taken as the signal from the primary nucleus, and  the charge of this nucleus was calculated as $Q=\sqrt{A\times cos{\theta}}$ where $\theta$ is the zenith angle. The final conclusion about the charge of the primary nucleus was done after a correction for the logarithmic increase of ionization with increasing energy: $Q(E_{d})= Q(100)/(1+0.012 \times log10(E_d/100))$, where $E_d$ is energy deposit in the calorimeter in GeV, and $Q(100)$ is the value of $Q$ at $E_d$=100 GeV. 

The charge resolution in the region of Z from 6 to 26 and for the energy deposition in the calorimeter above 50 GeV is shown in fig.1 (a,b). The peaks from abundant even nuclei are clearly seen, while elements of subiron group: Sc(Z=21), Ti(Z=22) and V(Z=23) are not resolved adequately. We show that by selecting the region of $21.5<Q<22.5$, we will exclude the contribution from the nearby abundant nuclei, while misidentification between titanium nuclei and neighboring nuclei of Sc and V should not distort the energy dependence of Ti/Fe because Sc and V  are also secondary. 

The silicon matrix was designed so that the pixels were arranged in four layers, and were partially overlapped. The area of overlap was about 15 $\%$ of the total matrix area. This allowed us to use results obtained by the double layer silicon matrix events  to determine  the  charge distribution for a nucleus  $Z$, resulting from the ionization loss fluctuations as well as from the variation  of the silicon pixel thickness. Fig.2(a) shows the distribution of the charge measured in the bottom layer pixel while carbon nuclei ($5.5<Q<6.5$) were selected in the top layer pixel. The form of the ionization loss is practically symmetric for the sufficiently high value of charge. However, the distribution has some tails due to the distribution of  thicknesses of the pixels. As a result, the total curve should be fit with a sum of two Gaussian curves: a narrow one for  ionization loss fluctuations for the bulk of the standard pixels, and  a  wider curve which describes the fluctuation of thickness for a small part of the non-standard pixels.

Selecting the proper range of ionization signal for a nucleus of charge Z in the top pixel, we  determined the shape of the charge line by the bottom pixel. The fit shows that the center of the narrow Gaussian is at $Z=5.94$ and the center of the wider Gaussian is at $Z=5.1$, while r.m.s. are 0.75 and 1.0 respectively. The ratio of the numer of nonstandard pixels to the number of standard ones is $15\%$. With the knowledge of these values, we can determine to what value of $Q$ the contribution of the given nucleus extends.  Fig. 2(b) shows the Q distribution of the calcium nuclei ($Z=20$) in the bottom pixel while Q value were selected in the range from 19.5 to 20.5 in the top pixel, and, similarly, fig.2(c) shows the distribution for iron nuclei  selected as $25.5 < Q < 26.5$. In this way it was determined that neither primary iron nucleus ($Z=26$) nor partially primary nuclei Ca ($Z=20$) or Cr ($Z=24$), could contribute significantly to the charge range from 21.5 to 22.5, which was selected for Ti nuclei. 
There are at least three more uncertainty which could distort the measured number of Ti nuclei. The first is influence of the albedo which could distort charge measuring in the silicon matrix. This problem was analized in \cite{albedo}, where we showed that albedo signals can not imitate charges with $Z>15$ at any incident particle energy. The second  is due to interaction of the primary cosmic ray flux in the silicon pixels or in the nearby material. This process was simulated with the FLUKA Monte Carlo codes for the detailed instrument model and it was found to be significant for light nuclei (see \cite{panov2}), but appears to be negligible for Ti nuclei. The third is connected with  the fragmantation of iron nuclei in the residual atmosphere ($\sim$ 5 g/cm$^2$). This correction was also determined with the simulation by FLUKA and is $1.66\%$ of the Fe flux at the top of the atmosphere and does not depend on energy. 
\begin{figure}[t]
 \begin{center}
 \includegraphics[width=20pc]{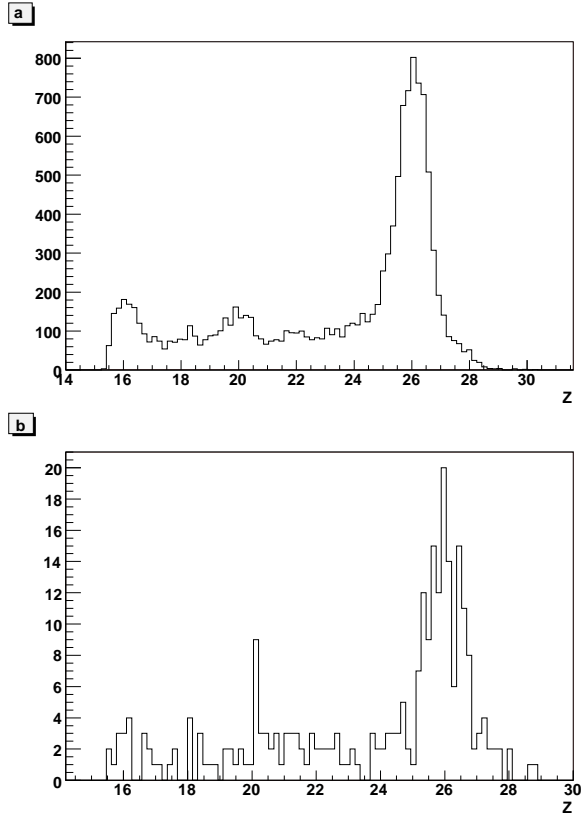}
  \end{center}
 \vspace{-1.0pc}
\caption{ Charge resolution for two regions of Ti energy: a) 5 GeV/n$<E_n<$80 GeV/n and b) $E_n>$80 GeV/n}
\end{figure}
\begin{figure}[t]
  \begin{center}
 \includegraphics[width=32pc]{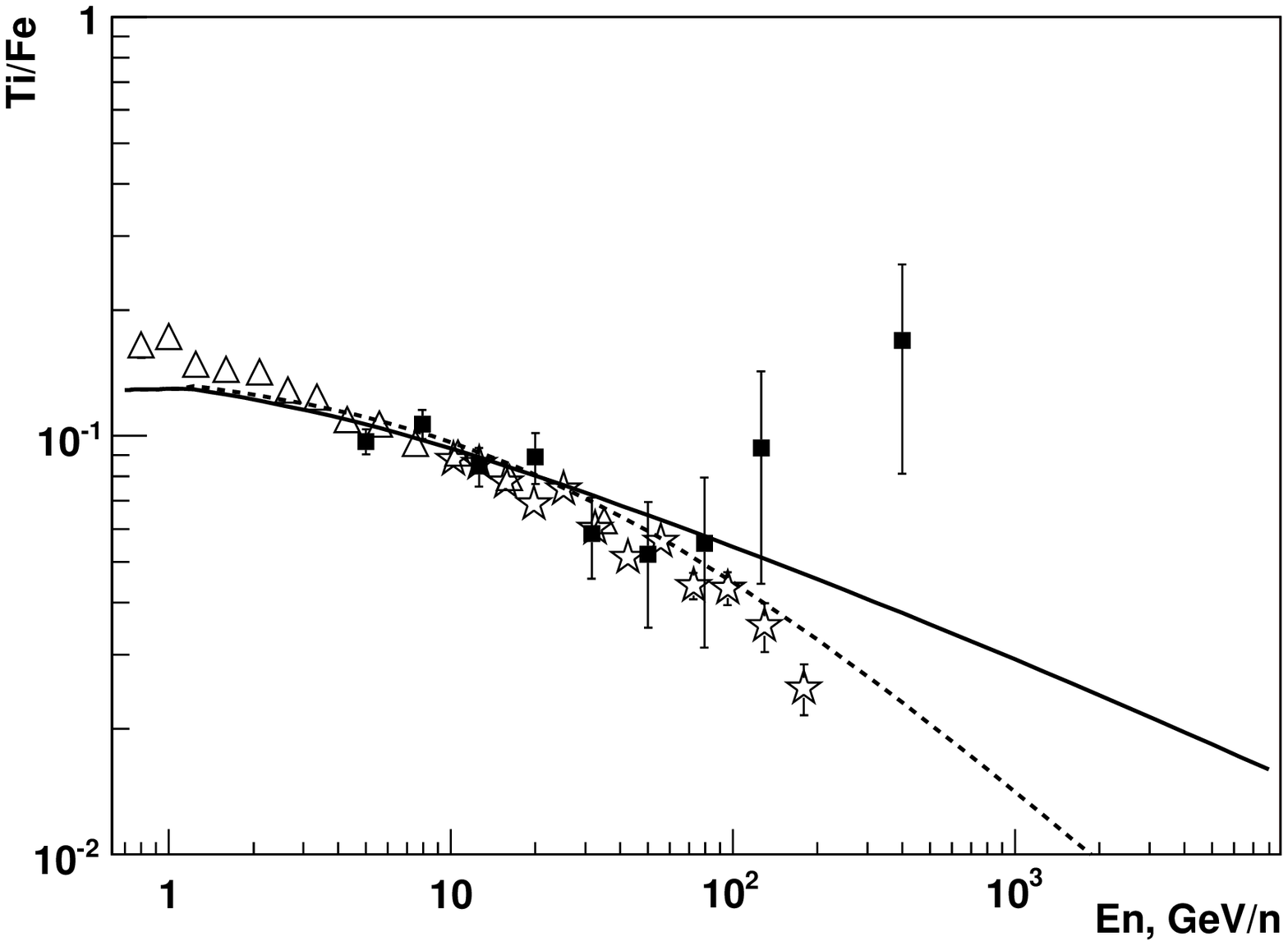}
  \end{center}
 \vspace{-1.0pc}
\caption{ Ratio of fluxes of Ti  and Fe.  Triangles:Engelmann et al.\cite{engelman}, stars:Vylet et al.\cite{Vylet}, squares: ATIC-2. Lower dashed line: Leaky-box model with $\lambda_{esc}\sim R^{-0.6}$, upper solid line: diffusion model of Osborn and Ptuskin} 
\end{figure}

The charge distributions for the charge region from 15 to 30 are shown in fig.3 for two energy regions of the titanium nuclei: a) from 5 GeV/n to 80 GeV/n and b) above 80 GeV/n. One can see that the charge resolution does not degradate as the energy increases. 
\section{Discussion}
Fig.4 shows the ratio of the fluxes of Ti and Fe from the ATIC-2 experiment along with the data of two experiments onboard the HEAO-3 satellite \cite{engelman}, \cite{Vylet}. Also shown are two model lines: for the Leaky-Box model with $\lambda_{esc} \sim R^{-0.6}$ (lower dashed line) \cite{engelman} and the diffusion model of Osborn and Ptuskin \cite{ptuskin} which assumed some reacceleration on interstallar magnetic irregularities (Kolmogorov model of magnetic irregularities) and rigidity dependence for the escape length in the form $R^{-0.33}$ (upper solid line). It is seen that the ATIC-2 data agree with the data of \cite{engelman} and \cite{Vylet}, and do not demonstrate a  flatter behaviour than the Leaky-Box model with $\lambda_{esc} \sim R^{-0.6}$ up to 100 GeV per nucleon. 

In the region above 100 GeV per nucleon, the data of \cite{Vylet} show the agreement with the dashed line, while the ATIC-2 experimental points are above both models. However, due to the limited statistics in the ATIC-2 dataset we can not do  a definite statement about  Ti/Fe ratio above 100 GeV.
\section{Acknowledgments}
This work was supported by the Russian Foundation for Basic Reseach (grant no. 05-02-16222) and the National Aeronautics and Space Administration of the United States (grants nos. NNG04WC12G, NNG04WC10G, and NNG04WC06G).
 
\end{document}